\documentclass[aps,twocolumn,prabib,amsmath,amssymb,superscriptaddress,notitlepage]{revtex4-1}
\usepackage{graphics}
\usepackage{bm}
\usepackage{dcolumn}
\usepackage{natbib}
\usepackage{epstopdf}
\usepackage{float}
\usepackage{graphicx}
\usepackage{epsfig}
\usepackage[pdfstartview=FitH]{hyperref}
\usepackage{color}
\usepackage{appendix}
\usepackage{ulem}

\newcommand{\rr}{{\bf r}}

\begin{document}
\title{Phase transitions of bosonic fractional quantum Hall effect in topological flat bands}
\author{Tian-Sheng Zeng}
\affiliation{Department of Physics, School of Physical Science and Technology, Xiamen University, Xiamen 361005, China}

\date{\today}
\begin{abstract}
We study the phase transitions of bosonic $\nu=1/2$ fractional quantum Hall (FQH) effect in different topological lattice models under the interplay of onsite periodic potential and Hubbard repulsion. Through exact diagonalization and density matrix renormalization group methods, we demonstrate that the many-body ground state undergoes a continuous phase transition between bosonic FQH liquid and a trivial (Mott) insulator induced by the periodic potential, characterized by the smooth crossover of energy and entanglement entropy. When the Hubbard repulsion decreases, we claim that this bosonic FQH liquid would turn into a superfluid state with a direct energy level crossing and a discontinuous leap of off-diagonal long-range order.
\end{abstract}
\maketitle
\section{Introduction}

In the past several decades the emergence of topological phases of matter beyond Landau's paradigm opens up an innovatory chapter in modern condensed matter physics~\cite{Wen2017}. One of the paramount topics is related to the phase transitions among these different topologically ordered phases such as fractional quantum Hall (FQH) effect~\cite{Laughlin1983} and spin liquid~\cite{Kalmeyer1987}. These topologically ordered phases, which are characterized by long-range entanglement without Landau's symmetry-breaking order parameters, usually host a well-defined topological invariant~\cite{Thouless1982}, e.g. fractional Hall conductance of the quantum Hall systems. Therefore the phase transition between a topologically ordered phase and a symmetry-broken phase is an intricate problem, which inspires much interest in its transition nature.

Indeed, a lot of theoretical studies on quantum Hall transitions between quantum Hall phases and topologically trivial phases, which have enlarged the domain of phase transition physics, are in bloom~\cite{Lam1984,Ludwig1994,Murphy1994,Morf1998,Evers2001,Evers2008}. Generally, there exist two possible scenarios for phase transitions separating a topologically ordered phase with another trivial phase by (i) a first-order transition such as a transition between two distinct ground states of an Ising quantum Hall ferromagnet~\cite{Piazza1999}, and (ii) a second-order phase transition, for instance Landau-forbidden transitions between a bosonic integer quantum Hall liquid and trivial insulator in two dimensions~\cite{Grover2013,Lu2014} which are also numerically examined in different topological models~\cite{Geraedts2017,Zeng2020}. Of particular interest, under the periodic chemical potential, a continuous transition between a fractional quantum Hall liquid at weak potentials and a Mott insulator at strong potentials is claimed in Refs.~\cite{Wen1993,Chen1993,Kol1993} on the basis of effective field theory. In contrast for disordered potentials, the localization/delocalization transition between plateaux in the fermionic integer quantum Hall system is shown to exhibit the universality class~\cite{Kivelson1992,Shahar1995}. Numerically, disorder-driven phase transition from a fractional quantum Hall liquid to an Anderson insulator is shown to be continuous~\cite{Liu2016}. In the presence of a spatial symmetry, it is argued that there may be a direct continuous transition between the bosonic $\nu=1/2$ FQH liquid and the bosonic superfluid~\cite{Barkeshli2014,Barkeshli2015}, whereas numerically the compressible-to-incompressible phase transitions in quantum Hall systems~\cite{Papic2009} are likely to be of first-order nature.

However in lattice models, prior major interests are focused on the creation or implementation of Laughlin-like bosonic FQH state~\cite{Lukin2005,Lukin2007,Kapit2010,Hormozi2012,Hayward2012,Zhu2016}, and much less knowledge has been gathered about the phase transition picture of quantum Hall physics against other phases. The topological flat bands without magnetic fields have become a basic infrastructure for studying the quantum Hall effect, with fractionalised topological phases predicted at partial fillings, dubbed ``fractional Chern insulators''~\cite{Sun2011,Neupert2011,Sheng2011,Tang2011,Wang2011,Regnault2011}. In cold atomic gases, topological Haldane-honeycomb lattice has been obtained in periodically driven optical lattice~\cite{Jotzu2014}, and topological Harper-Hofstadter Hamiltonian is obtained using either laser-assisted tunneling in neutral $^{87}$Rb atoms~\cite{Aidelsburger2013,Miyake2013,Aidelsburger2015,Greiner2017} or synthetic dimension in alkaline-earth-metal-like atoms with multiple internal degrees of freedom~\cite{Stuhl2015,Mancini2015}. Moreover in multi-layer systems, tunable Chern insulators under moir\'e superlattice potential have been achieved~\cite{Chen2019,Chen2020}. In Ref.~\cite{Spanton2017}, the existence of fractionalised interacting phases is experimentally confirmed, and the phase transitions between these quantized phases are mapped out. Such continuous transitions between different fractional quantum Hall states induced by periodic potentials have been intensively discussed regarding different Chern insulators in Ref.~\cite{Lee2018} with fascinating QED properties at the critical points. These experimental progresses hold promise for exploring exotic phase transition physics in fractional quantum Hall systems, such as the phase transition between FQH and other trivial phases in interacting Harper-Hofstadter model~\cite{Motruk2017,Rosson2019}.

In this work, we focus on the phase transition physics for softcore bosons in concrete topological lattice models at filling factor $\nu=1/2$ under the interplay of periodic potential and Hubbard repulsion. For strong Hubbard repulsion, Laughlin-like fractional quantum Hall effect is believed to emerge~\cite{Wang2011}. Upon the addition of periodic potential or the softening of Hubbard repulsion, we elucidate the phase transition nature between different competing phases through state-of-the-art density-matrix renormalization group (DMRG) and exact diagonalization (ED) simulations.

The rest of this paper is organized as follows. In Sec.~\ref{model}, we give a description of the Bose-Hubbard model Hamiltonian with periodic potential in different topological lattice models, such as $\pi$-flux checkerboard and Haldane-honeycomb lattices. In Sec.~\ref{mott}, we present the numerical results for the FQH-insulator transition induced by periodic potential. Further in Sec.~\ref{superfluid}, we present the numerical results for the FQH-superfluid transition when the Hubbard repulsion is decreased. Finally, in Sec.~\ref{summary}, we summarize our results and discuss the prospect of investigating such phase transitions in flat band systems.

\section{Models and Methods}\label{model}

\begin{figure}[b]
  \includegraphics[height=1.7in,width=3.4in]{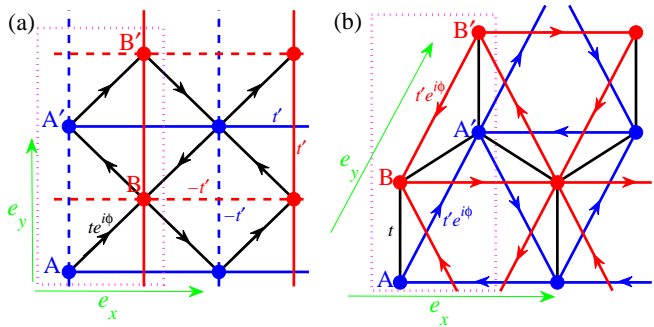}
  \caption{\label{lattice}(Color online) The schematic plot of (a) The $\pi$-flux checkerboard lattice model in Eq.~\ref{cb} and (b) the Haldane-honeycomb lattice model in Eq.~\ref{hc}. The arrow directions present the signs of the phases $\phi$ in the hopping terms.  Sublattice $A$ ($B$) is labeled by blue (red) filled circles. The arrow link shows the hopping direction carrying chiral flux $\phi_{\rr'\rr}$. For the checkerboard lattice, the next-nearest-neighbor hopping amplitudes are $t_{\rr,\rr'}'=\pm t'$ along the solid (dotted) lines. $e_{x,y}$ indicate the real-space lattice translational vectors.
  The magenta dotted box depicts the enlarged unit cell with periodic potentials $\mu_A=0,\mu_{A'}=\mu_B=\mu_{B'}=\mu$.}
\end{figure}

We begin with the following noninteracting tight-binding Hamiltonian in two typical topological lattice models, as shown in Figs.~\ref{lattice}(a) and~\ref{lattice}(b),
\begin{align}
  H_{CB}^0&=-t\!\sum_{\langle\rr,\rr'\rangle}\!\big[b_{\rr'}^{\dag}b_{\rr}\exp(i\phi_{\rr'\rr})+H.c.\big]\nonumber\\
  &-\!\sum_{\langle\langle\rr,\rr'\rangle\rangle}\!\!\! t_{\rr,\rr'}'b_{\rr'}^{\dag}b_{\rr}
  -t''\!\sum_{\langle\langle\langle\rr,\rr'\rangle\rangle\rangle}\!\!\!\! b_{\rr'}^{\dag}b_{\rr}+H.c.,\nonumber\\
  H_{HC}^0&=-t'\!\sum_{\langle\langle\rr,\rr'\rangle\rangle}[b_{\rr'}^{\dag}b_{\rr}\exp(i\phi_{\rr'\rr})+H.c.]\nonumber\\
  &-t\!\sum_{\langle\rr,\rr'\rangle}\!\! b_{\rr'}^{\dag}b_{\rr}
  -t''\!\sum_{\langle\langle\langle\rr,\rr'\rangle\rangle\rangle}\!\!\!\! b_{\rr'}^{\dag}b_{\rr}+H.c.,\nonumber
\end{align}
where $H_{CB}^0$ denotes the $\pi$-flux checkerboard (CB) lattice
and $H_{HC}^0$ the Haldane-honeycomb (HC) lattice. Here $b_{\rr}^{\dag}\quad(b_{\rr})$ is the particle creation (annihilation) operator at site $\rr$, $\langle\ldots\rangle$,$\langle\langle\ldots\rangle\rangle$ and $\langle\langle\langle\ldots\rangle\rangle\rangle$ denote the nearest-neighbor, the next-nearest-neighbor, and the next-next-nearest-neighbor pairs of sites, respectively. In the flat band parameters, we choose $t'=0.3t,t''=-0.2t,\phi=\pi/4$ for checkerboard lattice, while $t'=0.6t,t''=-0.58t,\phi=2\pi/5$ for honeycomb lattice, as in Refs.~\cite{Wang2011,Wang2012}.

Further we take the Bose-Hubbard repulsion as $ V_{int}=U/2\sum_{\rr}n_{\rr}(n_{\rr}-1)$, where $U$ is the onsite interaction strength and $n_{\rr}=b_{\rr}^{\dag}b_{\rr}$ is the particle number operator at site $\rr$. In what follows, we will numerically address the many-body ground states of interacting bosons in the presence of periodic potentials, and the full Hamiltonian is written as
\begin{align}
  H_{CB}&=H_{CB}^0+\frac{U}{2}\sum_{\rr}n_{\rr}(n_{\rr}-1)+\sum_{\rr}\mu_{\rr}n_{\rr},\label{cb}\\
  H_{HC}&=H_{HC}^0+\frac{U}{2}\sum_{\rr}n_{\rr}(n_{\rr}-1)+\sum_{\rr}\mu_{\rr}n_{\rr},\label{hc}
\end{align}
where periodic potential $\mu_{\rr}$ is chosen with commensurate period two: $\mu_{\rr}=0$ for $A$ sites
while $\mu_{\rr}=\mu$ for $B,A',B'$ sites within each unit cell, as shown in Figs.~\ref{lattice}(a) and~\ref{lattice}(b).

In the ED study, we study a finite system of $N_x\times N_y$ unit cells (the total number of sites is $N_s=2\times N_x\times N_y$). The total filling of the lowest Chern band is $\nu=N/(N_xN_y)$ with global $U(1)$-symmetry. With the translational symmetry, the energy states are labeled by the total momentum $K=(K_x,K_y)$ in units of $(2\pi/N_x,2\pi/N_y)$ in the Brillouin zone. While the ED calculations on the periodic lattice are limited to a small system, we exploit infinite DMRG for larger systems on cylinder geometry with open boundary conditions in the $x$-direction and periodic boundary conditions in the $y$-direction. We keep the bond dimension up to 3600 to obtain accurate results for different system sizes.

\section{Numerical Results for FQH-insulator Transition}\label{mott}

In this section, we present our numerical results for the transition between a FQH liquid and a Mott insulator induced by periodic potential $\mu$ at a given filling $\nu=1/2$ for bosons. For strong Hubbard repulsion the system is known to fall into the $\nu=1/2$ FQH phase at $\mu=0$, as demonstrated in Ref.~\cite{Wang2011}. When $\mu$ increases, the $\nu=1/2$ FQH would be overwhelmed by a trivial insulator where the particles are localized at strong $\mu/t\gg1$. In the following parts, we will discuss this phase transition from several aspects including ground state degenerate manifold and entanglement entropy.

\begin{figure}[t]
  \includegraphics[height=2.7in,width=3.4in]{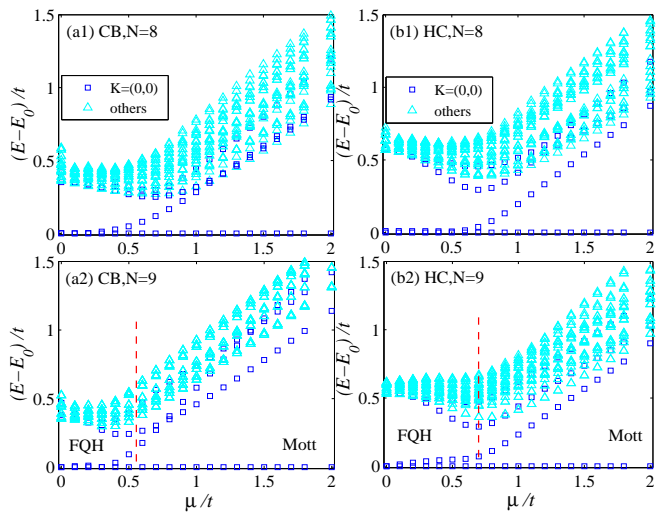}
  \caption{\label{energy}(Color online) Numerical ED results for the low energy spectrum of Bose-Hubbard system $\nu=1/2,U=\infty$ as a function of periodic potential $\mu$ on two typical topological lattices: (a) $\pi$-flux checkerboard and (b) Haldane-honeycomb lattices, respectively. In the presence of periodic potential $\mu$, the unit cell is doubled with four inequivalent lattice sites $A,B,A',B'$ and the two-fold degenerate FQH ground states fall into the same momentum sector.}
\end{figure}

\subsection{ED results}

\begin{figure}[t]
  \includegraphics[height=1.6in,width=3.4in]{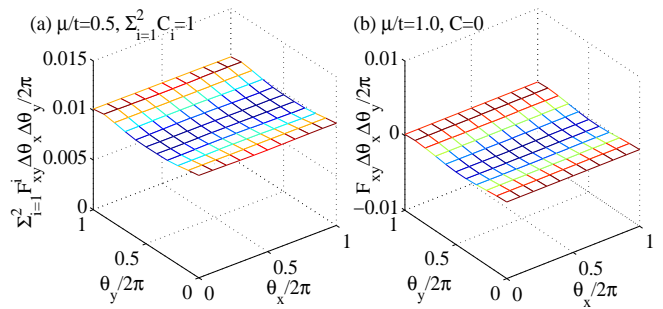}
  \caption{\label{berry}(Color online) Numerical ED results for Berry curvatures $F_{xy}\Delta\theta_{x}\Delta\theta_{y}/2\pi$ in hardcore bosonic systems $\nu=1/2,N_s=2\times4\times4$ for: (a) the $K=(0,0)$ two ground states at $\mu<\mu_c$; (b) the $K=(0,0)$ unique ground state at $\mu>\mu_c$.}
\end{figure}

We first present an ED study of the ground state properties in Eqs.~\ref{cb} and~\ref{hc} with hardcore limit $U/t=\infty$.
In Figs.~\ref{energy}(a) and~\ref{energy}(b), we plot the low-energy spectrum as a function of on-site periodic potential $\mu$ for various system sizes in different topological lattices.
For weak potentials, there always exist two nearly-degenerate ground states with a large gap separated from higher excited levels, which is the hallmark characteristic of $\nu=1/2$ FQH liquid. The question whether or not these ground states are FQH states against other competing trivial phases like charge-density-wave phases, can be answered by their topological invariants, i.~e. the Chern numbers. In the presence of periodic potential, these two-fold ground states are coupled with each other in the same momentum sector within the reduced Brillouin zone. With twisted boundary conditions~\cite{Niu1985,Thouless1989,Hatsugai2005}, $\psi(\rr+N_{x}\widehat{e}_x)=e^{i\theta_{x}}\psi(\rr)$,
$\psi(\rr+N_{y}\widehat{e}_y)=e^{i\theta_{y}}\psi(\rr)$ where $\theta_{x,y}$ are the twisted angle, the Chern number is given by $C=\int\int d\theta_{x}d\theta_{y}F_{xy}/2\pi$ where the Berry curvature is given by $F_{xy}=\mathbf{Im}(\langle{\partial_{\theta_x}\psi}|{\partial_{\theta_y}\psi}\rangle
-\langle{\partial_{\theta_y}\psi}{\partial_{\theta_x}\psi}\rangle)$. By calculating the smooth Berry curvatures shown in Fig.~\ref{berry}(a), we obtain a well-quantized total Chern number $\sum_{i=1}^2C_i=1$ for two gapped ground states at $\mu<\mu_c$. By tuning $\mu$ from weak to strong, the ground states do not undergo the level crossing with excited levels for different system sizes, signaling a continuous phase transition nature. Across a threshold value $\mu=\mu_c$, these two-fold ground states split and a unique ground state is left in the Mott insulator for $\mu>\mu_c$. We confirm that this gapped unique ground state hosts a zero Chern number $C=0$ at $\mu>\mu_c$, as indicated by the vanishing Berry curvatures in Fig.~\ref{berry}(b).

\begin{figure}[t]
  \includegraphics[height=1.65in,width=3.4in]{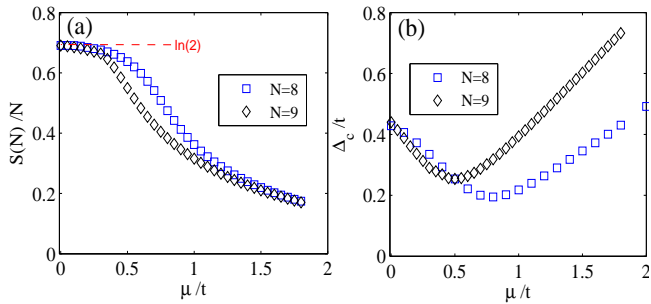}
  \caption{\label{gap}(Color online) Numerical ED results for the FQH-insulator transition at $\nu=1/2,U=\infty$ in the CB lattice as periodic potential $\mu$ is increased from weak to strong for different sizes: (a) single-particle entanglement entropy $S(N)$ and (b) charge-hole gap $\Delta_c$.}
\end{figure}

Next, we examine the change of single-particle entanglement entropy for interacting $N$-particle systems as a function of on-site periodic potential $\mu$. Here we diagonalize the reduced single-particle density matrix $\widehat{\rho}_{\rr,\rr'}=\langle\psi|b_{\rr}^{\dag}b_{\rr'}|\psi\rangle$ with $N_{s}\times N_{s}$ elements, and obtain single-particle eigenstates $\widehat{\rho}|\phi_{j}\rangle=n_{j}|\phi_{j}\rangle$ where $|\phi_{j}\rangle$ ($j=1,\ldots,N_{s}$) are the natural orbitals and $n_{j}$ ($n_1\geq\ldots\geq n_{N_{s}}$) are interpreted as occupations. The single-particle entanglement entropy is defined as
\begin{align}
  S(N)=-\sum_{j=1}^{N_s}n_{j}\ln n_{j}.
\end{align}
For $\nu=1/2$ FQH liquid, the particles uniformly occupy the lowest Chern band with the occupations $n_{j}\simeq 1/2$ for $j\leq N_s/2$ and $n_{j}\ll1$ for $j>N_s/2$. Thus $S(N)$ becomes a universal constant $N\times\ln2$. However, increasing periodic potential $\mu$ leads to the splitting of the lowest band into two subbands, and the particles tend to occupy the lower subband with $n_{j}\simeq 1$ for $j\leq N=N_s/4$ and $n_{j}\ll1$ for $j>N$, in order to minimize the total energy. For the Mott insulator, the $N$ particles are localized at certain $N$ sites, and $S(N)$ tends to zero for strong $\mu/t\gg1$. As shown in Fig.~\ref{gap}(a), the single-particle entanglement entropy $S(N)$ evolves smoothly from weak $\mu$ to strong $\mu$, which serves as another signature of continuous FQH-insulator transition.

Moreover we continue to discuss the insulating behavior against particle excitations across the FQH-insulator transition. As is well-known, both $\nu=1/2$ FQH liquid and Mott insulator are incompressible phases characterized by the presence of the charge-hole gap. Here we calculate the charge-hole gap $\Delta_c(\mu)=[E_{N+1}(\mu)+E_{N-1}(\mu)-2E_{N}(\mu)]/2$ where $E_{N}(\mu)$ is the ground energy for interacting $N$-particle systems. As shown in Fig.~\ref{gap}(b), $\Delta_c$ shrinks as $\mu$ increases from zero to $\mu_c$, implying the softening of $\nu=1/2$ FQH liquid, while it tends to dilate for strong $\mu>\mu_c$, which scales as $\Delta_c(\mu)\propto\mu$ when the particle excitation is controlled by the periodic potential. Nevertheless, $\Delta_c(\mu)$ hosts a nonzero minimum cusp at the threshold point $\mu=\mu_c$ for different system sizes, indicating the continuous transition. However, one should be careful to extract the excitation information in the thermodynamic limit. According to the construction of effective QED$_3$-Chern-Simons theory~\cite{Lee2018}, this critical point should be described by one Dirac fermion coupled to a gauge field and the gap closing should happen at certain high symmetry momentum point like the $\Gamma$ point $K=(0,0)$ in the Brillouin zone as indicated in Fig.~\ref{energy} for all system sizes. Thus we expect the excitation gap in Figs.~\ref{energy} and~\ref{gap}(b), due to finite size effect, would collapse at the critical point $\mu=\mu_c$ in the thermodynamic limit, which is beyond current scope of our computability.

\subsection{DMRG results}

Following the last section, we move on to discuss the continuous transition between $\nu=1/2$ FQH liquid and Mott insulator from the perspective of DMRG simulation. Here we exploit an unbiased DMRG approach for larger system sizes, using a cylindrical geometry up to a maximum width $N_y=8$. For topologically ordered phases like $\nu=1/2$ FQH liquid, they host nonlocal anyons with long-range entanglement in the ground state, reflected in the topological entanglement entropy. As pointed out in Refs.~\cite{Kitaev2006,Levin2006}, the entanglement entropy $S_L$ of a partitioned subsystem of a gapped two-dimensional system satisfies the volume area
\begin{align}
  S_L(\ell)=\alpha \ell-\gamma+\cdots,
\end{align}
where $\ell$ is the boundary length of the subsystem. The topological entanglement entropy $\gamma=\ln(D)$ is a universal constant with $D$ the total quantum dimension, i.e. $D=\sqrt{2}$ for Laughlin $\nu=1/2$ FQH liquid. In our infinite DMRG, we divide the cylinder into left and right parts along the $x$ direction~\cite{Jiang2012}, and calculate the entanglement entropy of the left part as $S_L(\ell)$ with the boundary length $\ell=2N_y$.

\begin{figure}[t]
  \includegraphics[height=2.65in,width=3.4in]{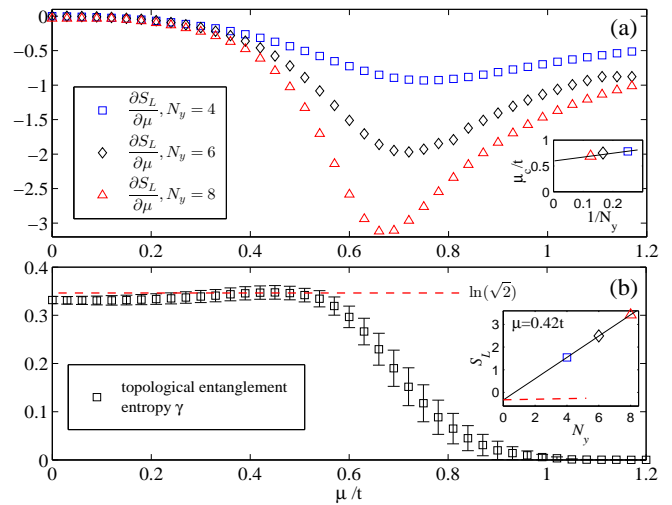}
  \caption{\label{dmrg}(Color online) Numerical DMRG results for the FQH-insulator transition at $\nu=1/2,U=\infty$ in the CB lattice as periodic potential $\mu$ is increased from weak to strong: (a) firs-order derivative of entanglement entropy and (b) topological entanglement entropy $\gamma$. The lattice geometry is taken with finite cylinder width $N_y$ and infinite cylinder length $N_x=\infty$. The upper inset plots the finite scaling of critical periodic potential $\mu_c$ as the cylinder width increases. The lower inset plots the typical extraction of topological entanglement entropy $\gamma$ using the finite area-law scaling of entanglement entropy at given potential $\mu=0.42t$.}
\end{figure}

We calculate the entanglement entropy $S_L$ for three different widths $N_y=4,6,8$, which varies smoothly as $\mu$ increases. In the Mott insulator, there is no fractionalization of anyons, and the topological entanglement entropy $\gamma=0$. As shown in Fig.~\ref{dmrg}(a), the first-order derivative of $S_L$ as function of $\mu$ for different cylinder widths exhibits a local minimum at $\mu=\mu_c$ which persists to a non-vanishing finite value as the cylinder width increases, indicating a continuous phase transition even in the thermodynamic limit~\cite{Liu2016,Chen2006,Hamma2008}.

Meanwhile, we scale $S_L$ as a function of $N_y$, and obtain the topological entanglement entropy $\gamma$ for a given $\mu$. As shown in Fig.~\ref{dmrg}(b), for $\mu<\mu_c$, $\gamma$ remains close to the theoretical value $\ln(\sqrt{2})$, consistent with the prediction of $\nu=1/2$ FQH liquid. However for $\mu>\mu_c$, $\gamma$ decreases monotonically down to zero as $\mu$ increases. Together with $S_L$, the analytic behavior of $\gamma$ demonstrates the continuous quantum phase transition from $\nu=1/2$ FQH liquid to a Mott insulator.

\section{Numerical Results for FQH-superfluid Transition}\label{superfluid}

In this section, we now turn to discuss the numerical results for the transition between a FQH liquid and bosonic superfluid induced by the softening of Hubbard repulsion $U$ at a given filling $\nu=1/2$ for bosons. For weak Hubbard repulsion $U/t\ll1$, the $\nu=1/2$ FQH liquid should be destroyed and the weakly interacting bosons would condense into the lowest single-particle kinetic orbital, where the system becomes a gapless superfluid. In the following parts, we will discuss quantum phase transition from several aspects including ground state degenerate manifold and the off-diagonal long-range order $\langle b_{\rr}^{\dag}b_{\rr'}\rangle$. Here to bypass the numerical difficulty, we take the maximum particle occupation per site $N_{max}=2$ for Bose-Hubbard model in Eqs.~\ref{cb} and~\ref{hc}, which is a very good approximation for the low lattice filling $N/N_s=1/4\ll1$.

\begin{figure}[t]
  \includegraphics[height=2.0in,width=3.4in]{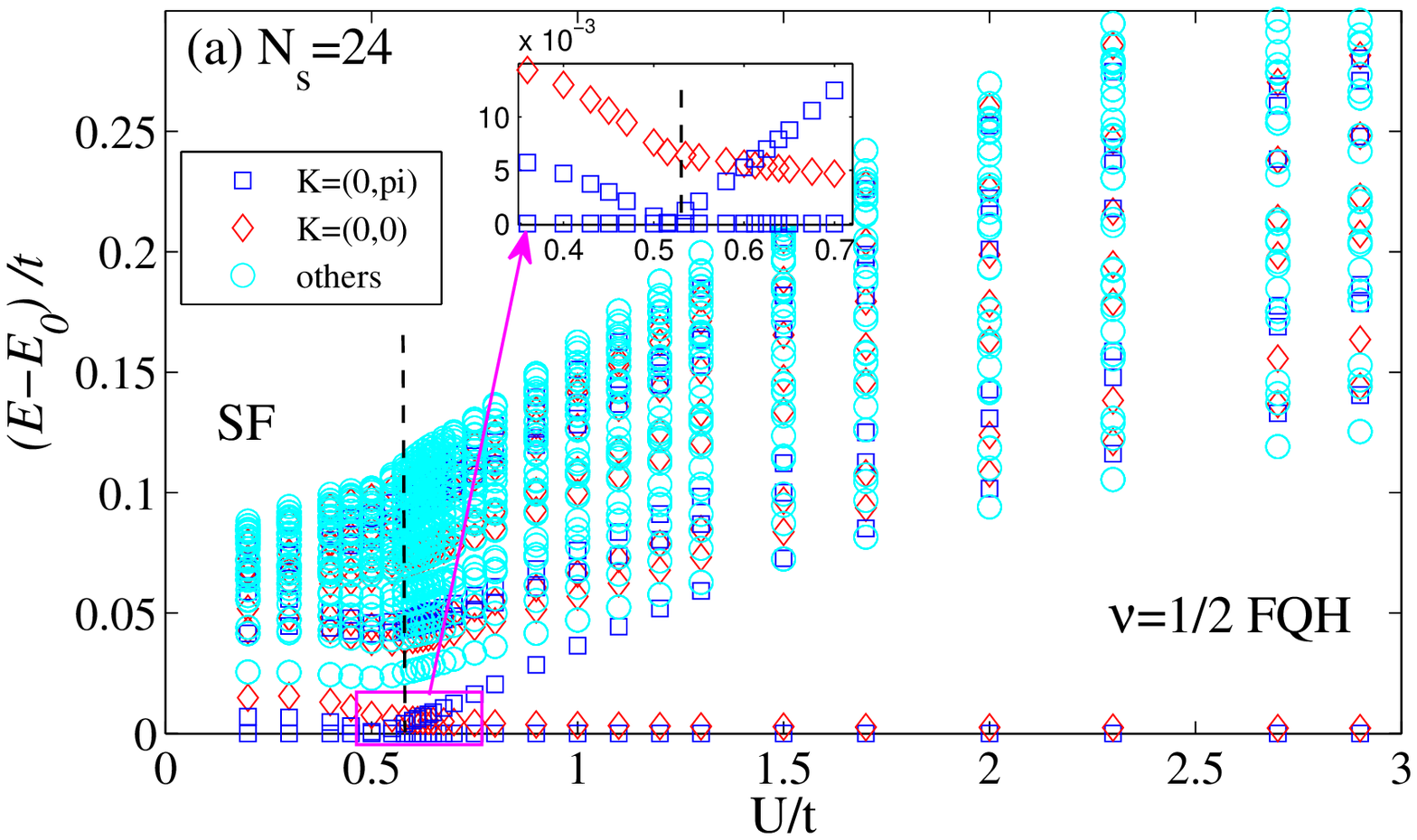}
  \includegraphics[height=2.0in,width=3.4in]{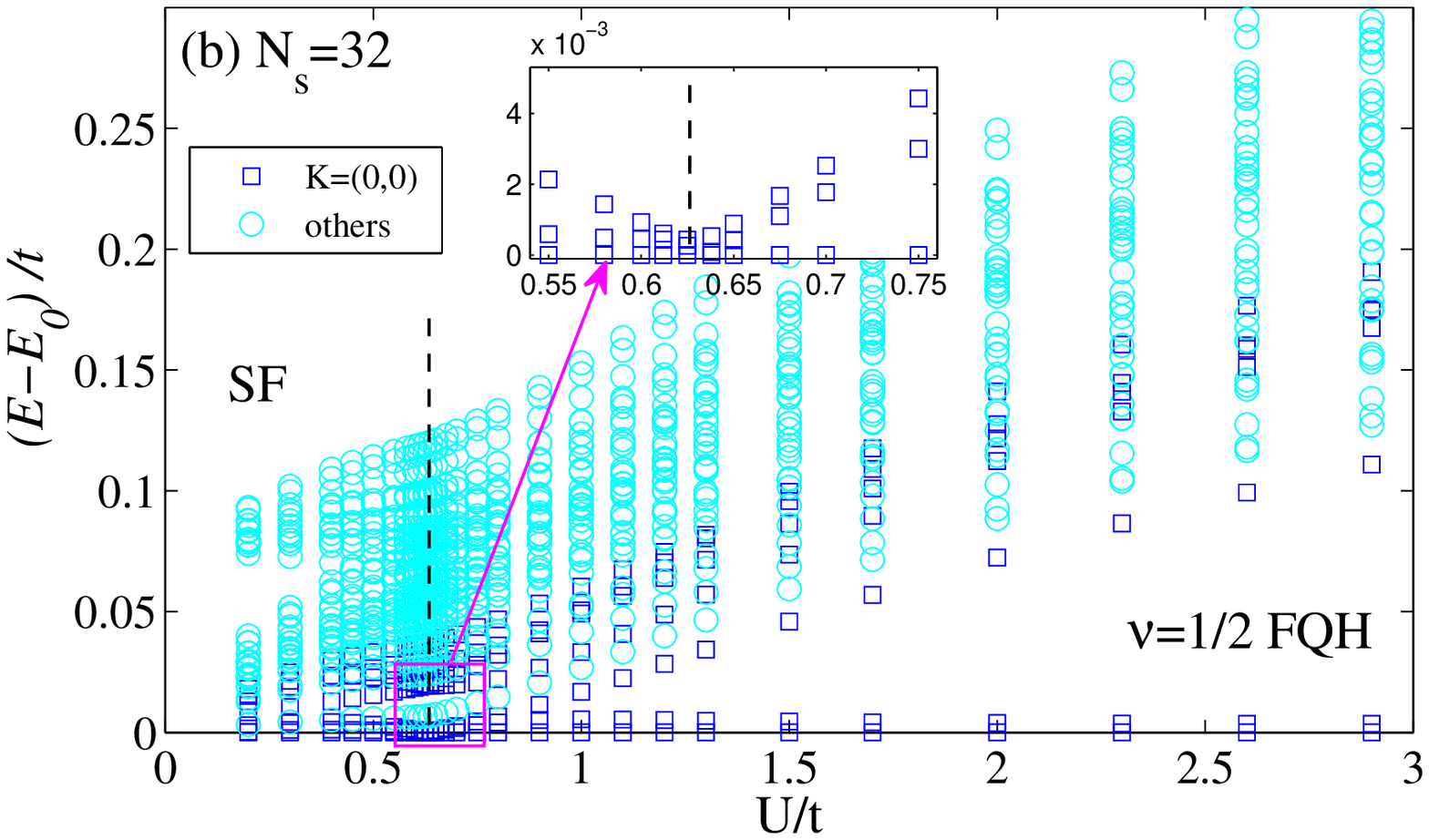}
  \caption{\label{sf}(Color online) Numerical ED results for the low energy spectrum of Bose-Hubbard system $\nu=1/2,\mu=0$ as a function of Hubbard repulsion $U$ on topological $\pi$-flux checkerboard lattice for different system sizes (a) $N_s=2\times3\times4$ and (b) $N_s=2\times4\times4$ respectively. The black dashed indicates the level-crossing point of the lowest ground state. The insets shows the zoom-in scan of the low energy spectrum near the transition point.}
\end{figure}

As indicated in Figs.~\ref{sf}(a) and~\ref{sf}(b) for different system sizes, we plot the low energy evolution at $\nu=1/2$ on topological $\pi$-flux checkerboard lattice as a function of Hubbard repulsion $U$. When $U$ decreases, the energy gap protecting the two-fold ground state degeneracy diminishes, and finally near the transition point $U\simeq U_c$ these two ground states undergo the direct level crossing with the third excited level, signaling a first-order phase transition nature. Moreover the momentum sector of the lowest ground state remains unchanged across the phase transition. Across the transition point $U<U_c$, there are many low-lying excited energy levels and the system enters into a bosonic superfluid. Similarly, we also confirm that the level crossing is observed on topological Haldane-honeycomb lattice with a smaller $U_c$, regardless of detailed lattice geometry.

\begin{figure}[t]
  \includegraphics[height=3.0in,width=3.4in]{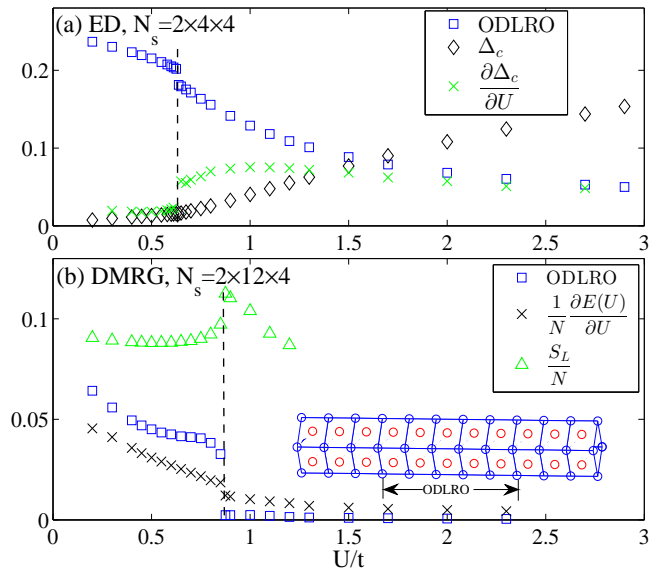}
  \caption{\label{odlro}(Color online) Numerical results for the FQH-superfluid transition at $\nu=1/2,\mu=0$ on topological $\pi$-flux checkerboard lattice as Hubbard repulsion $U$ is decreased from strong to weak for different system sizes: (a) $N_s=2\times4\times4$ in ED and (b) $N_s=2\times12\times4$ in DMRG. The black dashed line labels the discontinuous transition point $U_c$ of key physical quantities, including off-diagonal long-range order (ODLRO), first-derivative of ground state energy per particle $\partial E(U)/(N\partial U)$, and half-cylinder entanglement entropy $S_L$. The inset cartoon depicts the measurement of ODLRO in the middle regime of finite cylinder, in order to eliminate the open boundary effect at two ends (The blue (red) filled circles label Sublattice $A$ ($B$)). Here finite DMRG is used with the maximal bond dimension 2420. }
\end{figure}

In order to demonstrate the first-order transition, we investigate the change behavior of the off-diagonal long-range order (ODLRO) related to bosonic superfluid. For finite system sizes, we define ODLRO as $\langle b_{0}^{\dag}b_{s}\rangle$ where $s=|\rr|_{max}$ is the most remote distance relative to the original point $\rr=0$ in periodic lattice. As shown in Fig.~\ref{odlro}(a), when $U$ weakens, bosonic phase coherence is enhanced and a discontinuous jump of ODLRO appears at the transition point $U=U_c$, consistent with the level crossing in Fig.~\ref{sf}(b). Further, we calculate the charge-hole gap $\Delta_c(U)=[E_{N+1}(U)+E_{N-1}(U)-2E_{N}(U)]/2$ which serves as a characterization tool between incompressible and compressible liquids. $\Delta_c$ decreases monotonously in the FQH region as $U$ is tuned down to $U_c$, and then for $U<U_c$, $\Delta_c$ hosts a very small value, i.e. of the order $0.01t$ limited by our finite system size. Due to the level crossing, the nonanalytic behavior of $\Delta_c$ at $U=U_c$ is also consistent with the first-order transition.

For a larger finite cylinder system, we also confirm the first-order transition nature, as evidenced by multiple measures through DMRG simulation. As indicated in Fig.~\ref{odlro}(b), first, the first-order derivative of the ground state energy hosts a small jump near the transition point $U_c$, in support of the level-crossing picture in ED. Second, when the ODLRO is measured inside the cylinder, it manifests a discontinuous jump at $U=U_c$, as it is. Third, the half-cylinder entanglement entropy $S_L$ is also nonanalytic with a leap in the slope at $U=U_c$, in sharp contrast to that in the FQH-insulator transition. Note that in Figs.~\ref{sf} and~\ref{odlro}, as the system size increases, the transition point is shifted towards a slightly bigger value, implying the existence of phase transition driven by Hubbard repulsion in the thermodynamic limit.

Finally we remark that our complementary ED and DMRG calculations in various detailed lattice models give a contrary deduction to the conclusion drawn by Ref.~\cite{Barkeshli2014} where instead an effective continuous transition is derived. In the thermodynamic limit, it may be likely that the weakly first-order transition is smeared by other perturbations like disorder in realistic experimental environments, and a continuous transition may intervene since disorder would couple the two states in the same momentum sector with a level repulsion left.

\section{Summary and Discussions}\label{summary}

In summary, we have studied different competing phases under the interplay of interaction and periodic potential. We show that the continuous phase transition between bosonic $\nu=1/2$ FQH liquid and a featureless (Mott) insulator is induced by the periodic potential, characterized by the smooth crossover of ground state energy and entanglement entropy. In contrast as the Hubbard repulsion decreases, we find that this bosonic FQH liquid would undergo a weakly first-order transition into a gapless superfluid with a sudden change of off-diagonal long-range order. Our softcore bosonic flat band models would be generalized to a larger class of interacting Hamiltonian featuring $\nu=1/2$ FQH effect, such as interacting Harper-Hofstadter models, which is of sufficient feasibility to be realized for current experiments in cold atoms~\cite{Cooper2019}. We emphasize that, actually the phase transitions between $\nu=1/2$ FQH liquid and other competing phases tuned by periodic potential in interacting Harper-Hofstadter models~\cite{Motruk2017} are of the same nature as these in our flat band models, regardless of any details in lattice structure, demonstrating the universal physical picture in these phase transitions.

\begin{acknowledgements}
T.S.Z thanks D. N. Sheng for inspiring discussions and prior collaborations on fractional quantum Hall physics in topological flat band models. This work is supported by the start-up funding from Xiamen University. T.S.Z also acknowledges partial support from the National Natural Science Foundation of China (NSFC) under the Grant No. 12074320.
\end{acknowledgements}

\end{document}